\documentclass[aps,prl,showpacs,twocolumn,amsmath,amssymb]{revtex4-1}

\usepackage{graphicx}
\usepackage{dcolumn}
\usepackage{bm}
\usepackage{color}
\usepackage{upgreek}
\usepackage{ulem}

\begin{document}

\title{Mesoscopic Rydberg-blockaded ensembles in the superatom regime and beyond} 
\author{T.M. Weber}
\author{M. H\"oning}
\author{T. Niederpr\"um}
\author{T. Manthey}
\author{O. Thomas}
\author{V. Guarrera$^1$}
\author{M. Fleischhauer}
\author{G. Barontini$^2$}
\author{H. Ott}
\email{ott@physik.uni-kl.de}
\affiliation{Research Center OPTIMAS, Technische Universit\"at Kaiserslautern, 67663 Kaiserslautern, Germany}
\affiliation{$^1$ Present address: LNE-SYRTE, Observatoire de Paris, CNRS, UPMC, 61 avenue
de l'Observatoire, 75014 Paris, France
 \\
$^2$ Present address: Laboratoire Kastler Brossel, ENS, UPMC-Paris 6, CNRS, 24
rue Lhomond, 75005 Paris, France
}



\maketitle

\begin{figure*}[t]
\begin{center}
\includegraphics[width=\textwidth]{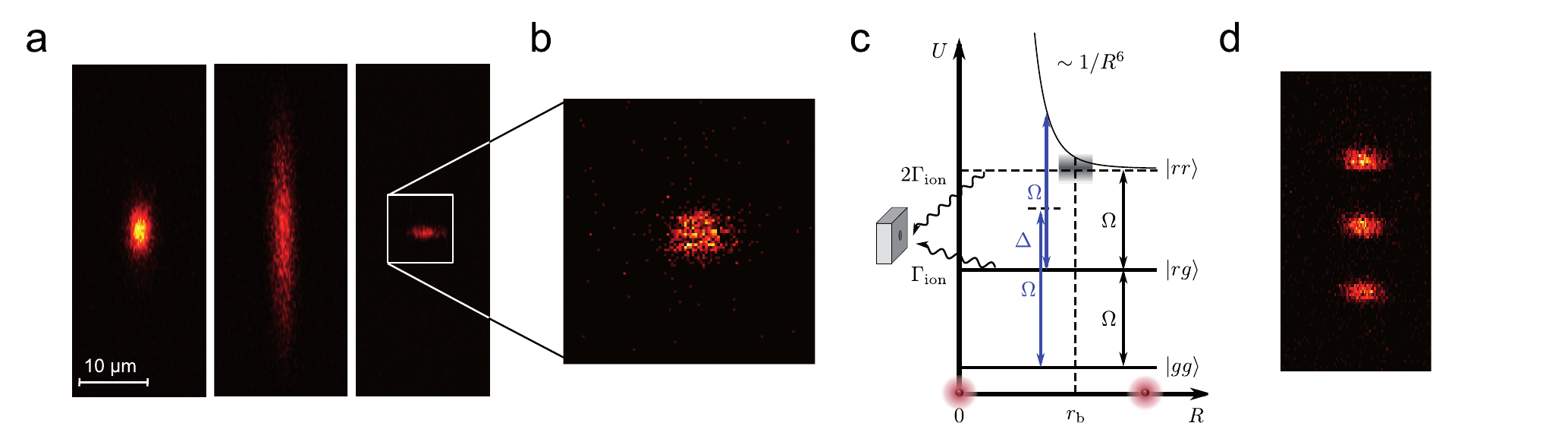}
\end{center}
\caption{\textbf{Preparation of superatoms and excitation level scheme.} (a) Starting from a Bose-Einstein condensate in an optical dipole trap 
with superimposed optical lattice (left) the atoms are radially compressed (middle) and removed from all but a few lattice sites with help of a 
focused electron beam (right). (b) The resulting almost spherical atomic sample contains between 100 and 500 atoms at a temperature of 
$3-4$\,$\upmu$K and dimensions smaller than 3\,$\upmu$m (4$\sigma$ of a Gaussian fit) in each direction. The axial size of this atomic sample can be varied arbitrarily. (c) The superatom is excited with a single photon transition ($\lambda=297\,$nm) into the $\left|51p_{3/2}\right\rangle$-state. Effective three-level scheme, comprising the 
ground state, the singly excited state and the doubly excited state, whose energy has an $r^{-6}$ dependency due to the van-der-Waals 
interaction. Black (blue) arrows denote (off-)resonant excitation and the blockade radius $r_B$ is denoted by the excitation linewidth (grey shaded area). The decay due to ionization and the ion detection are also indicated. (d) A one-dimensional array of superatoms, demonstrating the scalability of our approach.}
\label{fig1}
\end{figure*}
 
\textbf{Controlling strongly interacting many-body systems enables the creation of tailored quantum matter, with properties transcending those based 
solely on single particle physics. Atomic ensembles which are optically driven to a Rydberg state provide many examples of this, such as atom-atom entanglement \cite{Gaetan2009,Urban2009}, many-body Rabi oscillations \cite{Dudin2012}, strong photon-photon interaction \cite{Peyronel2012} 
and spatial pair correlations \cite{Schauss2012}. In its most basic form, Rydberg quantum matter consists of an isolated ensemble of strongly 
interacting atoms spatially confined to the blockade volume - a so-called \textit{superatom}. Here we demonstrate the controlled creation and characterization 
of an isolated mesoscopic superatom by means of accurate density engineering and excitation to Rydberg $p$-states. Its variable size
allows to investigate the transition from effective two-level physics for strong confinement to many-body phenomena in extended systems.
By monitoring continuous laser-induced ionization we observe a strongly anti-bunched ion emission under blockade conditions and extremely bunched 
ion emission under off-resonant excitation. 
Our experimental setup enables \textit{in vivo} measurements of the superatom, yielding insight into both excitation statistics and dynamics. We anticipate straightforward applications in quantum optics and quantum information as well as future experiments on many-body physics.}\\

Rydberg superatoms combine single and many-body quantum effects in a unique way and have been proposed as fundamental building blocks for 
quantum simulation and quantum information \cite{Saffman2010}. Due to the phenomenon of Rydberg blockade \cite{Lukin2001}, the ensemble collectively forms a system with only two levels of excitation. Provided a range of interaction larger than the sample size, the presence of one excitation shifts all other atoms out of resonance and therefore only one excitation can be created at a time. Changing the size or the driving conditions revives the underlying many-body nature and the presence of several excited atoms with pronounced correlations becomes possible. This tunability and the possibility of multiple usage within a single experimental sequence make superatoms a promising complement to single-atom-based quantum technology. It is therefore important to understand the significance of the superatom concept, the 
implications of its finite spatial extent and its many-body level structure. We here investigate the latter by measuring the mean Rydberg excitation 
as well as its time-resolved two-particle correlations in an optically excited, mesoscopic superatom
for varying excitation strength and under resonant and non-resonant conditions, revealing very different excitation dynamics. 

The realization of superatom-based quantum systems requires the implementation of arbitrary arrangements of isolated mesoscopic atomic ensembles
in a scalable way. We here prepare an individual superatom by carefully shaping the density distribution of a Bose-Einstein condensate of $^{87}$Rb 
atoms. We first load the condensate into a one-dimensional optical lattice with a spacing of 532\,nm, in order to suppress the axial movement of the atoms. 
We subsequently compress the atomic sample in the radial direction to reduce its size below the blockade radius and empty all but three (or more) lattice 
sites using a focused electron beam \cite{Gericke2008,Wuertz2009,Barontini2013} (Fig.\,1a, Methods). The atom number within the ensemble $N$ can 
be adjusted between 100 and 500 at a temperature of $T=(3.5\pm0.5)\,\upmu$K and the typical size of the sample is $\leq3\,\upmu$m in diameter (Fig.\,1b). Our 
preparation scheme is readily scalable to arrays of superatoms (Fig.\,1d). 

After preparation we excite the atomic ensemble with a single photon transition from the $\left|5s_{1/2}\right\rangle$-ground state to the $\left|51p_{3/2},m_j=3/2\right\rangle$-Rydberg state at a wavelength of 297\,nm with coupling strength $\Omega$. The single photon transition circumvents scattering from any intermediate state and therefore allows for a long-time exposure. The key observable is the string of ions produced by excited atoms that are photoionized by the trap laser \cite{Anderson2013} (with ionization rate $\Gamma_{\mathrm{ion}}=(45\pm5)\,$kHz). Specifically, we detect the initial peak ion rate as well as the temporal pair correlation function $g^{(2)}(\tau)$, extracted from the time-resolved ion signal (see Methods). The ions continuously emitted from the ensemble lead to a slow decay of the superatom on a timescale between a few milliseconds (see inset Fig.\,2) and seconds. Eventually almost 100\,\% of the constituent atoms are converted into ions, of which we detect $(40\pm8)$\,\%. \\
%
\begin{figure}[h!]
\begin{center}
\includegraphics[width=0.85\columnwidth]{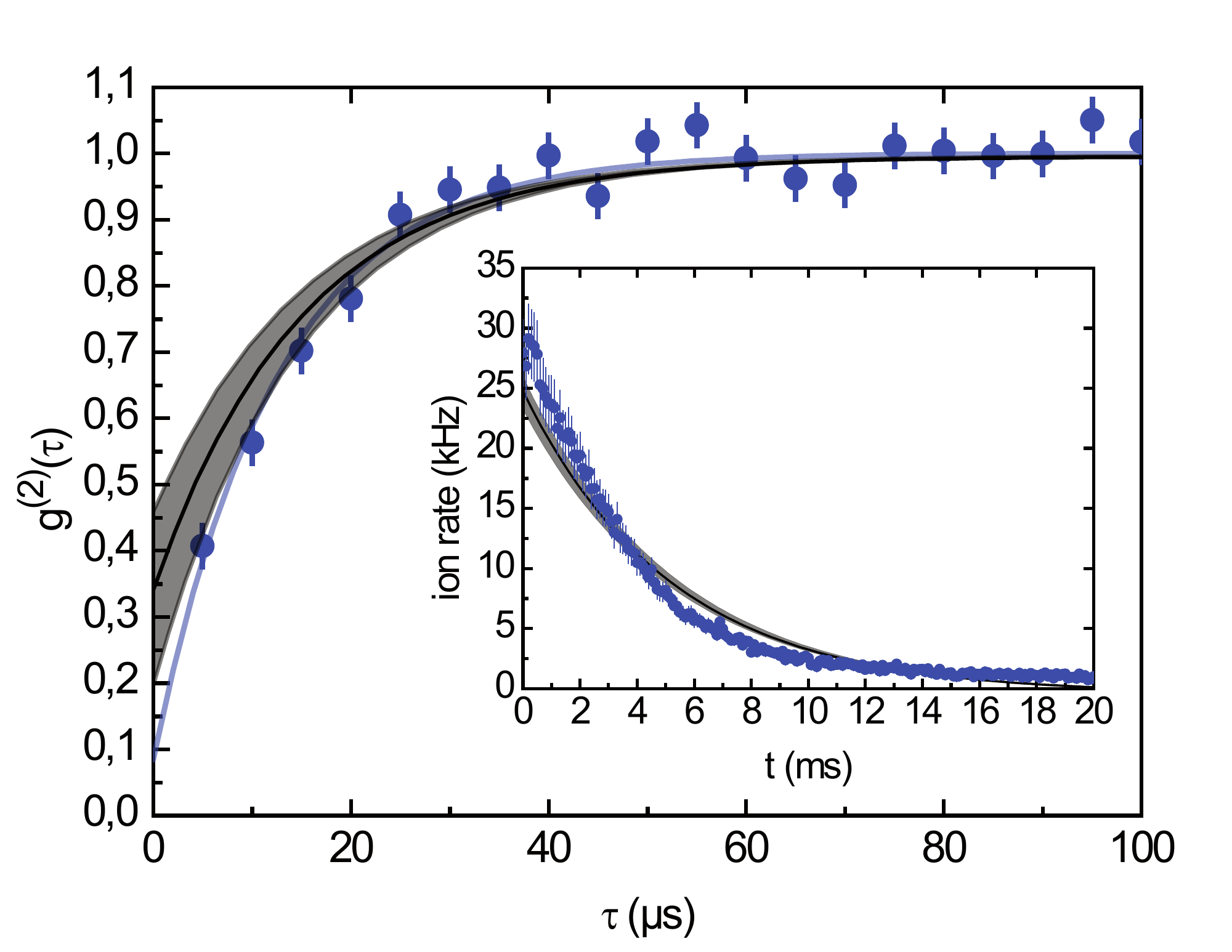}
\end{center}
\caption{\textbf{Resonant excitation of an isolated superatom.} (a) Second order temporal correlation function of ions emitted from the resonantly excited superatom ($\Omega/2\pi = 6\,$kHz, blue circles). Data points are fitted with an exponential function (blue curve), revealing a value of $g^{(2)}(0)=$0.08$\pm$0.06. The black line and the grey shaded area result from a rate model (see Methods and Supplementary information). The inset shows the data as well as the model curves of the absolute ion rate as a function of time. The integral below both curves is fixed by the total number of produced ions. The bars denote the statistical errors from 1500 experimental runs.}
\label{fig2}
\end{figure} 
%

For weak resonant driving the superatom mimics an effective two-level system where the excitation of more than one atom is suppressed due to blockade.
This can be clearly observed in the experiment: Fig.\,2) shows pronounced anti-bunching in good agreement with a theoretical rate model (Methods and Supplementary information). We extrapolate a value of $g^{(2)}(0)=0.08\pm0.06$, taking into account an uncorrelated background signal (see Methods). The background constitutes 10 - 15 \,\% of the signal and originates from atoms which are not removed during the preparation of the superatom. The anti-bunching amplitude stays constant during the gradual decay of the superatom, which thus acts as a continuously operating single ion source \cite{Ates2013}. The strong suppression of collective oscillations, indicated by the purely exponential shape of $g^{(2)}(\tau)$, shows that the system is in the overdamped regime, where the coherent coupling is overcome by decoherence from laser linewidth, thermal motion of atoms and residual field fluctuations as well as intrinsic dephasing mechanisms \cite{Honer2011}.
%
\begin{figure*}[t]
\begin{center}
\includegraphics[width=1\textwidth]{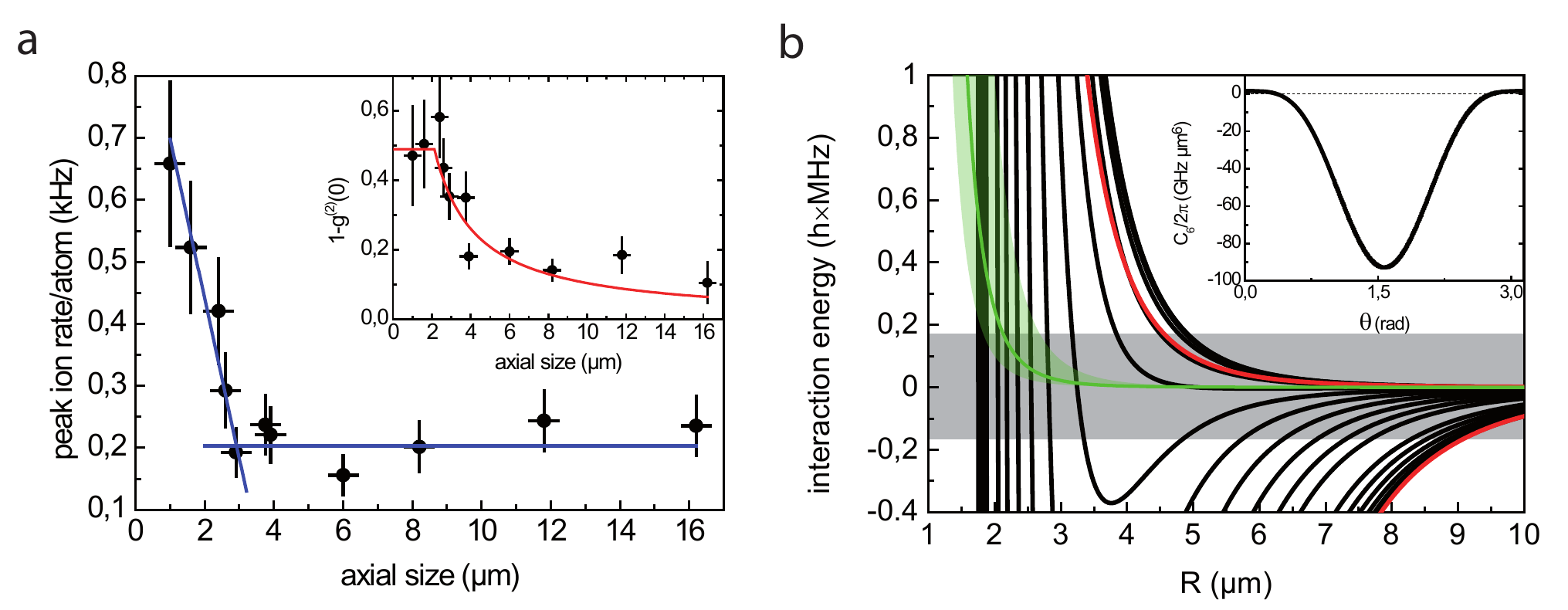}
\end{center}
\caption{\textbf{Blockade radius measurement and theoretical calculations.} (a) Initially emitted ion rate per atom for different axial sizes $l$ of the atomic sample (black circles). The density is kept constant for all sample sizes. We determine the effective blockade radius from the intersection of partial linear fits (blue curves). The inset shows the corresponding dependence of $g^{(2)}(0)$ on $l$ (black circles). The data are compared to a compound theory (red curve) of a hard shell model (for $l\leq r_B$) and a statistical decay $1-g^{(2)}(0)=(1-g^{(2)}_{l<r_B}(0))\cdot r_B/l$ (for $l>r_B$). The bars indicate the error of the fit for the peak ion rate, the sample size and the initial $g^{(2)}$-amplitude. (b) Potential curves of the asymptotic $\left|51p_{3/2,3/2},51p_{3/2,3/2}\right\rangle$ pair state in dependence on the inter-atomic distance $R$ as obtained from diagonalization of the interaction Hamiltonian for equidistant orientation angles from $\theta=0$ to $\theta=\pi/2$ (black curves). Red curves indicate the $C_6$-potentials for $\theta=0$ and $\theta=\pi/2$. The grey, shaded area denotes the region of resonant excitation for a decoherence rate $\Gamma_\mathrm{d}/2\pi = 340\,$kHz (see Methods). The green curve and area show the result of the effective rate model with an effective van-der-Waals coefficient $C_6^{\rm eff}/2\pi={16}^{+48}_{-10}$\,MHz$\cdot \upmu$m$^6$. The inset displays the angular dependency of $C_6$ of the asymptotic $\left|51p_{3/2,3/2},51p_{3/2,3/2}\right\rangle$ pair state on $\theta$ obtained from second order perturbation calculations.}
\label{fig3}
\end{figure*} 
%
The ability to adjust the size of the atomic sample allows for a continuous transition from the superatom limit to a many-body system, where blockade conditions break down. First of all this can be used to determine the blockade radius.
In Fig.\,3a the initial ion rate per atom is shown for increasing axial size $l$ of the sample, keeping the ground state atom density constant. Under blockade conditions the ion rate first decreases as more and more atoms contribute to the same signal. However, above a critical spatial extent, which we identify with the blockade radius, the ion rate remains constant. The corresponding, independently measured, anti-bunching signal (inset in Fig.\,3a) leads to a compatible value of the blockade radius of (2.7$\pm$0.8)\,$\upmu$m. At first glance it is quite surprising that we observe blockade at all when we resonantly excite to a $p$-state. In a $p$-state the van der  Waals interaction is strongly angular dependent (Fig.\,3b). For a small intervall of $\theta$ the coefficient $C_6$ is vanishingly small, potentially leading to a breakdown of the overall blockade. Thus to understand the observed blockade effect we need to go beyond the standard $C_6$ asymptotics. Fig.\,3b shows the interaction potential curves for the asymptotic $\left|51p_{3/2,3/2},51p_{3/2,3/2}\right\rangle$ pair state, obtained by diagonalization of the interaction Hamiltonian (see Methods). As a result of an avoided crossing with the asymptotic $\left|51p_{3/2,1/2},51p_{3/2,1/2}\right\rangle$ pair state, which is energetically separated by an external magnetic field of 35\,G, the potential curves which have a negative $C_6$ coefficient for large distances bend into a repulsive interaction for smaller distances. Thus, the interaction potential for all angles becomes repulsive, enabling an overall blockade. 

In order to describe the complex interaction potential structure in an effective but simple way, we assume an isotropic repulsive interaction and solve the many-body problem within an approximate rate equation model with a $C_6^{\rm eff}$ coefficient as the only free parameter (see Methods). We chose a van-der-Waals coefficient of $C_6^{\rm eff}/2\pi={16}^{+48}_{-10}$\,MHz$\cdot \upmu$m$^6$ to compare the model to the experimental results. The resulting potential curves are indicated by the green shaded area in Fig.\,3b. Throughout this paper, we apply the effective model with these parameters to our data and find good agreement over a large range of laser intensities and detunings.  \\

Our mesoscopic superatom permits inter-atomic distances $R$ of up to $\approx 3\,\upmu$m. As a consequence, when the excitation is off-resonant the blockade conditions can be tuned into an anti-blockade \cite{Amthor2010} and pronounced bunching of the ion emission can be observed (Fig.\,4a). We find bunching values of up to $g^{(2)}(0)=61\pm8$ for large detunings (Fig.\,4a). This behaviour can be understood from the full level structure of the mesoscopic superatom, including excited states with more than one excitation and is well captured by our rate model: While the transition to the first collective Rydberg state is out of resonance, subsequent transitions into doubly excited states are shifted into resonance (Fig.\,1c) leading to a cascaded excitation process. 
%
\begin{figure*}[t]
\begin{center}
\includegraphics[width=\textwidth]{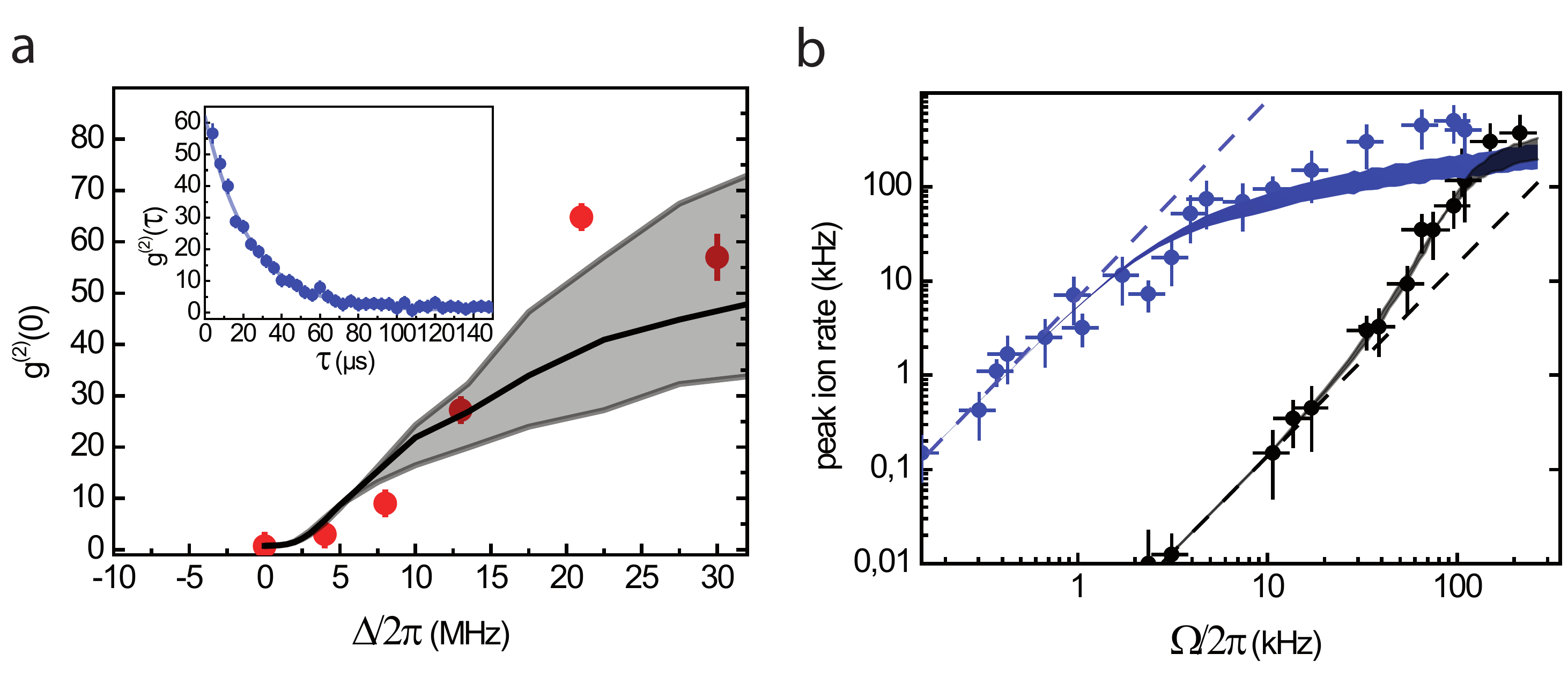}
\end{center}
\caption{\textbf{Off-resonant excitation and saturation of a mesoscopic superatom} (a) Pair correlation function $g^{(2)}(0)$ for different detuning (red circles, bars indicate the error of the fitted initial amplitude), compared to the model results (grey). Inset: $g^{(2)}(\tau)$ for a detuning of $\Delta/2\pi$ = 21\,MHz and $\Omega/2\pi$ = 48\,kHz (blue circles, bars indicate the statistical error from 600 experimental runs, the solid line is a fit with an exponential decay). (b) Dependence of the initial rate of ions emitted from the superatom on the coupling strength $\Omega/2\pi$ for resonant excitation (blue circles) and off-resonant excitation with blue detuning of $\Delta/2\pi$ = 4\,MHz (black circles) of an ensemble of 125 atoms. Data points are compared to the respective results of the rate model (shaded areas). Dashed lines continue the respective initial quadratic dependence. The vertical bars denote the error of the fit of the ion decay curve, the horizontal bars indicate the intensity fluctuations and drifts of the excitation laser.}
\label{fig4}
\end{figure*} 
%

The transition of the mesoscopic superatom from an effective two-level system to a complex many-level system is also reflected in its saturation behaviour. In Fig.\,4b, we plot the initial ion rate in dependence on the Rabi frequency for resonant and off-resonant ($\Delta/2\pi=4$\,MHz) excitation through three orders of magnitude of the experimental parameter $\Omega$. The excitation probability on resonance initially grows quadratically and starts to saturate around an ion rate corresponding to one excitation present. Driving the superatom more strongly, the blockade radius is reduced and above the saturation threshold more excitations can be created, resulting in an increasing initial ion rate, however with a smaller slope. Thus, the blockade is overcome after saturation has been reached. For off-resonant excitation the signal shows again a quadratic initial slope at a reduced absolute value, but enters a region where the slope is steeper than quadratic, showing a strong enhancement of excitations. For strong enough driving, the resonant and off-resonant excitation eventually reach a comparable level. This happens, when the collective coupling strength $\sqrt{N}\Omega$ becomes comparable with the detuning of 4\,MHz and the difference in the first excitation step for the resonant and off-resonant case disappears. The rate equation model reproduces all experimental findings despite the major simplifications made. Only for the largest Rabi frequencies in Fig.\,4b, the model underestimates the excitation rate. Here, the ensemble Rabi frequency $\sqrt{N}\Omega$ is larger than the decoherence rate and coherent many-body dynamics might become visible. However, a comparison of the rate equation model with a fully quantum-mechanical treatment leads to almost identical predictions for the initial ion rate and $g^{(2)}(0)$ (Supplementary information). In order to observe coherent dynamics, besides the condition $\Omega/\Gamma_\mathrm{d} >1$ also the blockade condition must be satisfied at the same time. The experiment is thus currently limited to incoherent dynamics. Reducing decoherence is one avenue towards coherent superatom dynamics. An alternative is the excitation to higher Rydberg $n$ levels, which increases the interaction.

\begin{figure}[t]
\begin{center}
\includegraphics[width=\columnwidth]{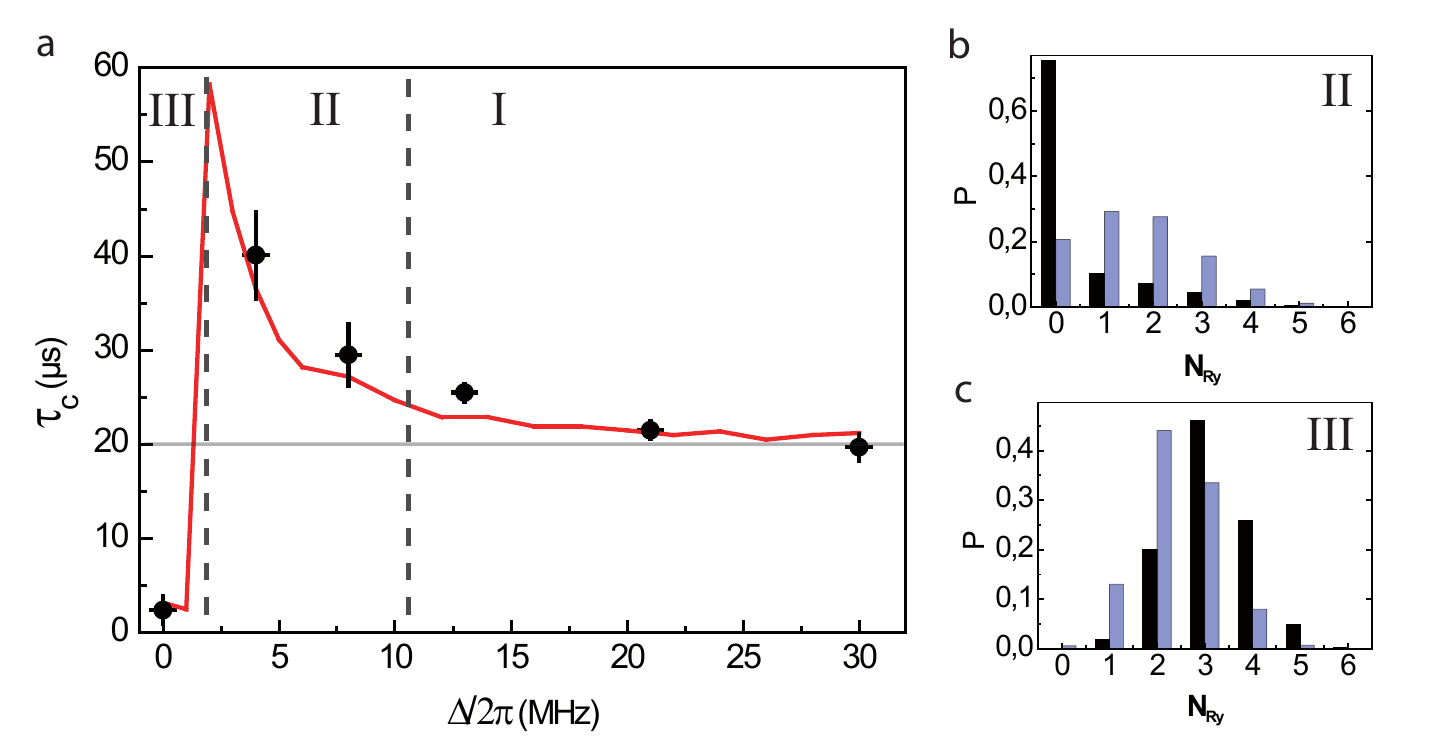}
\end{center}
\caption{\textbf{Dynamics of the pair correlation function} (a) Correlation time $\tau_c$ of the pair correlation function in dependence on the detuning (same parameters as in Fig.\,4a). The black points are the experimental data, the red line is the theoretical prediction. The correlation times are determined by an exponential fit (see Fig.\,4a). The bars indicate the error of the fit. The horizontal grey line indicates the lifetime of a Rydberg excitation. Three physical regimes can be distinguished (see text). (b) Calculated steady-state probability distribution of number of excitations (black bars) and corresponding non-equilibrium distribution after the detection of one ion (blue bars) for regime (II). The temporal evolution back to the steady-state is reflected in the correlation time of the pair correlation function. (c) Corresponding distributions for regime (III).}
\label{fig5}
\end{figure}

The temporal correlation function $g^{(2)}(\tau)$ also provides insight into the many-body dynamics of the superatom. Fig.\,5a shows the correlation times $\tau_c$ of $g^{(2)}(\tau)$ in dependence of the detuning. Three different physical regimes can be identified: (I) For large detunings, the atoms spend most of the time in the ground state with small probabilities for single and double excitations. The detection of an ion projects the density matrix onto states with one excitation less. Only doubly excited states emit a second ion and contribute to $g^{(2)}(\tau)$. The correlation time of $g^{(2)}(\tau)$ is thus simply given by the lifetime of the Rydberg excitation. (II) For smaller detunings we observe a dramatic slow-down of the relaxation dynamics. In this regime strongly correlated Rydberg aggregates form \cite{Schempp2014,Arimondo2013}. An ionization event projects the system onto a state with increased weight on aggregates of few excited atoms (Fig.\,5b). The relaxation to the steady-state is set by the lifetime of these aggregates, which exceed that of regime (I), as several atoms have to decay. (III) On resonance, where antibunching occurs, $g^{(2)}(\tau)$ reflects the excitation dynamics after the emission of an ion. The timescale is below $5\ \mathrm{\upmu s}$, shorter than all relevant single particle timescales, and is therefore a signature of the collectively enhanced excitation rate of the superatom. The corresponding probability distributions are shown in Fig.\,5c. Our analysis shows that the temporal pair correlation function is a powerful tool to characterize many-body dynamics and can be used in the future to characterize quantum phases of driven dissipative systems \cite{Adams2013}.\\\\

Single superatoms based on collective Rydberg excitations bear great potential for applications in quantum optics. They can be used to build high fidelity photon absorbers \cite{Honer2011} and deterministic ion sources \cite{ Ates2013}. An interaction between multiple superatoms can be realized choosing a F\"orster resonance, which features a long-range $R^{-3}$ dipole-dipole interaction. The increased interaction then enables us to switch to the coherent collective excitation regime, allowing for deterministic state manipulation. Strings of superatoms (see Fig.\,1d) are an ideal system for the investigation of energy transfer mechanisms \cite{Westermann2006, Rost2011} and one-dimensional spin systems \cite{Lesanovsky2012}. Our approach can be straightforwardly extended to arbitrary patterns of superatoms in two-dimensional lattice systems \cite{Gericke2008,Wuertz2009}. Such quantum systems can then be a resource for further investigations of isotropic 
and anisotropic long-range interactions \cite{Glaetzle2014}, for the quantum simulation of open spin systems \cite{Diehl2008,Hoening2014}, Bell state measurements \cite{Moebius2013,Wuester2013} or interferometric applications \cite{Wei2011,Gil2014}.\\ 

\subsubsection*{METHODS}

\textbf{Superatom preparation and laser excitation}\\

We start with a Bose-Einstein condensate of about 1700 rubidium atoms in a crossed optical dipole trap at a wavelength of $\lambda=1064\,$nm. The final trap frequencies are $2\pi\times$(260/85/270)\,Hz. A one-dimensional optical lattice is then superimposed by linearly ramping up a retro-reflected laser beam along the weak axis of the trap and the axial motion is frozen out. A focused electron beam ($(250\pm100)$\,nm radius, $20$\,nA beam current), which is also used to image the sample, removes the atoms from selected areas \cite{Gericke2008,Wuertz2009,Barontini2013}. In this way, we prepare samples of several hundred atoms with $\leq3\,\upmu$m diameter at a temperature of $(3.5\pm1)\,\upmu$K. After the preparation sequence, a fraction of 10 - 16\,\% of the atoms dwell in the outer regions of the trap.\\
The superatom is directly excited to a Rydberg state by illuminating it with an UV-laser beam at $297\,$nm with a waist of $100\,\upmu$m and laser power up to $160\,$mW. The light is produced by frequency doubling a stabilized dye laser (Mattisse-DR) in a heated CLBO-crystal installed in a Pound-Drewer-Hall stabilized bow-tie cavity. The frequency of the dye laser can be tuned via an offset-locked reference laser, resulting in a relative uncertainty of the UV frequency of $\pm0.5\,$MHz. The linewidth of the excitation light is estimated from dye laser control parameters to less than $200\,$kHz and the power noise is below 10\,\%.\\ 

\textbf{Electric fields, ion detection and signal processing}\\

The atomic sample is surrounded by quadruply segmented copper rings of 40\,mm diameter at a distance of 25\,mm, embodying an octupole electrodes configuration \cite{Manthey2014}. Applying corresponding voltages residual electric fields in the chamber are compensated in all directions apart from a remaining permanent vertical component of $E_0\leq$0.25\,V/cm that is used to extract produced ions towards the ion optics below. The ion optics guide the ions into a dynode multiplier (ETP 14553). The signal pulses are further processed with a temporal resolution of 100\,ns.\\

\textbf{Temporal correlation function}\\

We numerically calculate the second order temporal correlation function of the ion signal $I(t)$
\begin{equation}
g^{(2)}(\tau)=\frac{\left\langle I(t)I(t+\tau)\right\rangle}{\left\langle I(t)\right\rangle\left\langle I(t+\tau)\right\rangle}
\label{}
\end{equation}
where $\left\langle I(t)I(t+\tau)\right\rangle$ is calculated as the averaged product of counts (0 or 1) in two bins of distance $\tau$ and normalization $\left\langle I(t)\right\rangle\left\langle I(t+\tau)\right\rangle$ is given by the averaged ion rate. Several effects affect the data evaluation and have to be taken into account. Artefacts from detector ringing occur for time separations of less then 400\,ns. Coulomb repulsion between the ions during time of flight produces additional correlations on a timescale which depends on the extraction field. For our parameters, they never occur on timescales longer than 2\,$\upmu$s. We therefore discard all data points with $\tau\leq2\,\upmu$s. The background atoms (fraction $r$), contribute an uncorrelated signal to the ion emission of the superatom. This leads to a reduction of the measured amplitude $g^{(2)}_{meas}$ compared to the bare signal from the superatom $g^{(2)}_{real}$ which we correct for: $g^{(2)}_{real}(0)=(g^{(2)}_{meas}(0)-1)/(1-r)^2+1$. Note 
that $g^{(2)}(\tau)$ is independent of the detector efficiency.\\

\textbf{Calculation of the interaction potential}\\

The potential curves are calculated by diagonalization of the dipole-dipole interaction Hamiltonian in the presence of an electric and magnetic field. 
The basis set consists of all pair states which are closer than 15\,GHz in energy to the initial $\left|51p_{3/2,3/2},51p_{3/2,3/2}\right\rangle$ pair state. We consider all possible combinations of $s$-, $p$- and $d$- states including all Zeeman levels.\\

\textbf{Effective rate model}\\

The superatom dynamics is described by a Lindblad equation $\dot{\rho}=i(\sum_{j=1}^{N_\mathrm{atoms}} [\mathcal{H}_j,\rho]+[\mathcal{H}_\mathrm{int},\rho])+\sum_\nu \frac{1}{2}(2 L_\nu \rho L_\nu^\dagger-\{L_\nu^\dagger L_\nu,\rho\})$, where $\mathcal{H}_j=(\frac{\Omega}{2} |r\rangle\langle g|_j+h.c.)+\Delta |r\rangle\langle r|_j$ and 
$L_\nu$ are the jump operators for ionization ($\Gamma_\mathrm{ion}$), spontaneous decay into low-lying states which are not ionized ($\Gamma_\mathrm{sp}$) and decoherence ($\Gamma_\mathrm{d}$). The decoherence rate represents the cumulative effect of laser linewidth, thermal atomic motion, fluctuating electric fields and intrinsic dephasing mechanisms \cite{Honer2011}. The rates of ionization $(\Gamma_\mathrm{ion}=45\text{ kHz})$ and internal spontaneous decay $(\Gamma_\mathrm{sp}=5\text{ kHz})$ are known from independent measurements and we extract the excitation rate from the saturation measurements shown in Fig.\,4. For weak driving the ion signal is independent of the interaction term $\mathcal{H}_\mathrm{int}$ and given by $I_{\Omega\ll \gamma}=\frac{\Gamma_\mathrm{ion}N_\mathrm{atoms}}{\Gamma_\mathrm{ion}+\Gamma_\mathrm{sp}}\Omega^2\gamma/(\gamma^2+4\Delta^2)$, with $\gamma=\Gamma_\text{d}+\Gamma_\text{ion}+\Gamma_\text{sp}$, and $N_\mathrm{atoms}$ being the number of atoms within the superatom. Fits 
to the saturation measurements at $\Delta=0$ and $\Delta/(2\pi)=4\,\text{MHz}$ yield the relation between laser intensity and $\Omega$ and the decoherence rate $\Gamma_\text{d}/(2\pi)\approx 140\,\text{kHz}$ and $\Gamma_\text{d}/(2\pi)\approx 340\,\text{kHz}$, for two different sets of parameters used in the experiment. 

Beyond the regime of weak driving we describe the superatom by classical rate equations, which is justified by the large decoherence rate present in our setup and validated in previous studies of strongly interacting Rydberg systems using such methods \cite{Petrosyan2013, Ates2006, Schempp2014}. The system of rate equations describes dynamics in classical configuration space, where individual states are connected by single atom transitions at excitation rate $(P_i)$ and deexcitation rate $(D_i)$. These rates depend on the effective detuning of the atom $\delta_i=\Delta+\sum_{j}\frac{C_6^{\rm eff}}{|r_i-r_j|^6}|r\rangle\langle r|_j$ through $P_i=\frac{\Omega^2 \gamma}{4\delta_i^2+\gamma^2}$ and $D_i=\frac{\Omega^2 \gamma}{4\delta_i^2+\gamma^2}+\Gamma_\mathrm{ion}+\Gamma_\mathrm{sp}$. The set of many body rate equations is solved by stochastic sampling of trajectories. Simulations take into account the spatial distribution of atoms as measured in the experiment by averaging over many realizations. Atomic 
motion is not included within our description and the intricate $p$-state interaction is approximated with an effective, isotropic van-der-Waals potential $C_6^{\rm eff}/R^6$.

\bibliography{bibi}

\textbf{Acknowledgements}
\begin{acknowledgements}
We acknowledge financial support by the DFG within the SFB/TRR 49. V. G. and G. B. were supported by Marie Curie Intra-European Fellowships.\\\\ 
\end{acknowledgements}
\textbf{Author Contributions\\}
T. W., T. M., T. N., G. B., and H. O. designed and set up the apparatus.
G.B. and H. O. conceived the experiment.
T. W., T. N., O. T., T. M., and G. B. performed the experiment.
H. O. supervised the experiment.
T. W. analyzed the data and prepared the manuscript.
M. H. and M.F. developed the theoretical model.
M. H. made the numerical simulations.
M. F. supervised the numerical simulations.
All authors contributed to the data interpretation and manuscript preparation.

\end{document}